\documentclass[%
 aip,
rsi,%
 amsmath,amssymb,
 reprint,%
]{revtex4-1}

\usepackage{graphicx}% Include figure files
\usepackage{dcolumn}% Align table columns on decimal point
\usepackage{bm}% bold math
\begin{document}

\preprint{AIP/123-QED}

\title{Physical insights into the operation of a 1-nm gate length transistor based on MoS$_2$ with metallic carbon nanotube gate}

\author{Marta Perucchini}
 \affiliation{Dipartimento di Ingegneria dell'Informazione, Universit\`a di Pisa, Via G. Caruso 16, 56122 Pisa, Italy
}%Lines break automatically or can be forced with \\
\author{Enrique G. Marin}
\affiliation{Dipartimento di Ingegneria dell'Informazione, Universit\`a di Pisa, Via G. Caruso 16, 56122 Pisa, Italy
}%Lines break automatically or can be forced with \\ 
\author{Damiano Marian}
\affiliation{Dipartimento di Ingegneria dell'Informazione, Universit\`a di Pisa, Via G. Caruso 16, 56122 Pisa, Italy
}%Lines break automatically or can be forced with \\
\author{Giuseppe Iannaccone}
\affiliation{Dipartimento di Ingegneria dell'Informazione, Universit\`a di Pisa, Via G. Caruso 16, 56122 Pisa, Italy
}%Lines break automatically or can be forced with \\
\author{Gianluca Fiori}
\email{gfiori@mercurio.iet.unipi.it}
\affiliation{Dipartimento di Ingegneria dell'Informazione, Universit\`a di Pisa, Via G. Caruso 16, 56122 Pisa, Italy
}%Lines break automatically or can be forced with \\
%Corresponding author. Email:


\begin{abstract}
Low-dimensional materials such as layered semiconductors or carbon nanotubes (CNTs) have been attracting increasing attention in the last decades due to their inherent scaling properties, which become fundamental to sustain the scaling in electronic devices. Inspired by recent experimental results (S.B. Desai, S.R. Madhvapathy, A.B. Sachid, J.P. Llinas, Q. Wang, G.H. Ahn, G. Pitner, M.J. Kim, J. Bokor, C. Hu, H.-S. P. Wong, and A. Javey, Science 354, 99 (2016)), in this work we examined the ultimate performance of of MoS$_2$-channel Field Effect Transistors with 1-nm gate length by means of quantum transport simulations based on Poisson equation and Non-equilibrium Green's function formalism. We considered uniformly scaled devices, with channel lengths ranging from 5 to 20 nm controlled by a cylindrical gate with a 1-nm diameter, as would be required in realistic integrated circuits. Moreover, we also evaluated the effect of the finite density of states of a carbon nanotube gate on the loss of device performance. We noticed that the sub-threshold swing for all short-channel structures was greater than the ideal limit of thermionic devices and we attributed this to the presence of tunneling currents and gate-drain interactions. We tailored the transistor architecture in order to improve the gate control. We concluded that the limited CNT-channel capacitive coupling poses severe limitations on the operation and thus exploitation of the device.
%
%Valid PACS numbers may be entered using the \verb+\pacs{#1}+ command.
\end{abstract}

\keywords{2-D materials, Non-Equilibrium Green Functions, Field Effect Transistor, ballistic transport, carbon nanotube}%Use showkeys class option if keyword
\maketitle

%\begin{quotation}
%The ``lead paragraph'' is encapsulated with the \LaTeX\ 
%\verb+quotation+ environment and is formatted as a single paragraph before the first section heading. 
%(The \verb+quotation+ environment reverts to its usual meaning after the first sectioning command.) 
%Note that numbered references are allowed in the lead paragraph.
%
%The lead paragraph will only be found in an article being prepared for the journal \textit{Chaos}.
%\end{quotation}

\section{Introduction}
A recent experimental paper has proposed an intriguing possibility to scale down transistor size in the few-nm regime, consisting in the use of a metallic carbon nanotube (CNT) as the gate electrode \cite{Desai2016} of a field-effect transistor (FET) with a monolayer MoS$_2$ channel. Two-dimensional (2D) crystal semiconductors such as MoS$_2$ have been considered as interesting candidates for substituting silicon as channel material at the very end of the semiconductor roadmap (i.e. beyond the so-called 5-nm semiconductor technology node \cite{ITRS,Schwierz2015a,Fiori2014a,Javey2014mos2,Collaert2015,Radisavljevic2011,moore:1965}). In fact, differently from bulk semiconductors, the natural atomic confinement of the electrons in such layered structures allow a uniform control of the gate over the channel, thus reducing the so-called short-channel effects (SCE) which strongly hinder the performance of ultra-scaled devices: \cite{Chhowalla2016} the threshold voltage roll-off, the increase of tunneling currents and the drain-induced-barrier lowering (DIBL)\cite{Khanna2016,Tharun16}. Desai et al. \cite{Desai2016} suggest the use of one-dimensional (1D) structures to fulfill the function of the transistor gate as a way to overcome lithography limitations, at least in a laboratory environment. A similar approach, has been also pursued by Cao et al. \cite{Cao2016}, who demonstrated the use of ultra-thin synthesized cylindrical nanowires with a metallic core conformally coated with an high-$\kappa$ oxide \cite{Houssa2006}. However, both experimental works consider an extremely long channel which is of course not suitable for large-scale industrial applications. In \cite{Desai2016}, the source-to-drain distance is of the order of hundreds of nanometers leading to an  almost ideal sub threshold swing ($\sim$~65~mV per decade). In addition, metallic CNTs have a low density of states that could lead to degraded electrostatics in the case of short channels. In this work, we investigated the CNT-gate FET concept in order to understand the ultimate performance while scaling the device  down to ultra short channel lengths.

\begin{figure}[t]
	%\vskip-3.5ex
	\centering
	\includegraphics[width=1.0\columnwidth]{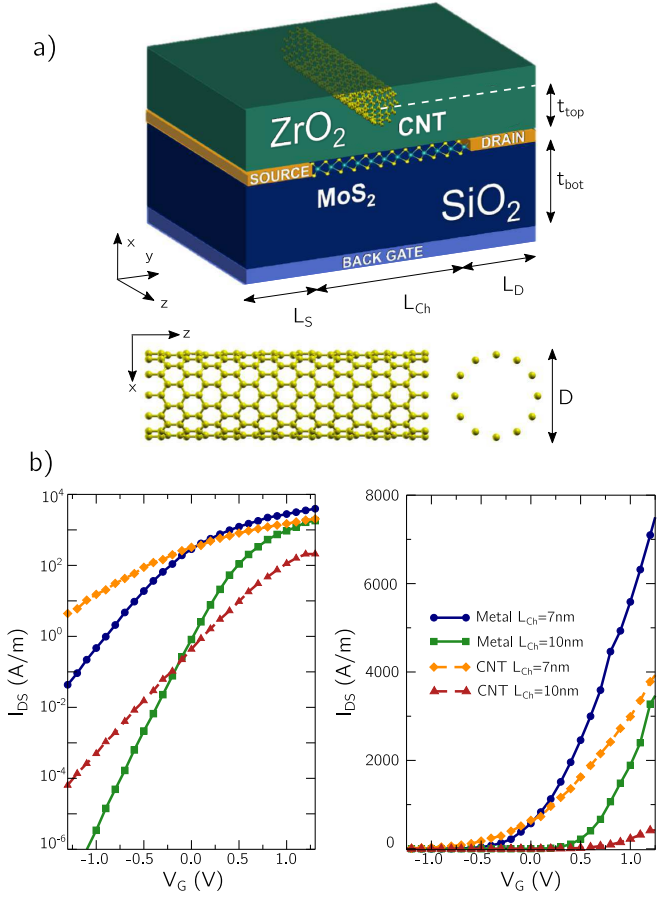}
	%\vskip-2ex
	\caption{(a) Top panel: sketch of the device structure: the top-gate is a single-walled metallic CNT with diameter $D$, separated from the channel by $t_{\text{top}}$ of ZrO$_2$. Source and drain contacts have lengths $L_{\text{S/D}}$, with an applied biases 0 and $V_{\text{DS}}$ respectively. The back-gate voltage is $V_{\text{BG}}$ and the bottom oxide (SiO$_2$) has a thickness $t_{\text{bot}}$. A single-layer of MoS$_2$ is used as channel, with length $L_{\text{Ch}}$. Bottom panel: lateral and front view of $(12,0)$ zig-zag CNT. (b) $I\!-\!V$ characteristics in semi-logarithmic (left panel) and linear scale (right panel) for devices with CNT or metallic gates with $D$=1nm and $L_{\text{Ch}}$=7nm and 10nm.}
	\label{Structure}
\end{figure}

\section{Models and Methods}
Aiming to determine the operation of the CNT-gated device shown in Fig.\ref{Structure}a,
we performed fully quantum transport simulations by self-consistently solving Poisson and Schr\"{o}dinger equations with open boundary conditions at room temperature (300K) \cite{Datta2000a} by means of the NanoTCAD ViDES simulation environment \cite{Fiori2006, Marin18,NanoTCAD}.
Starting from the experimental geometry presented by Desai et al. \cite{Desai2016}, we proposed  a double gated transistor with a MoS$_2$ monolayer channel of length $L_{\text{Ch}}$ as shown in Fig.\ref{Structure}a. 
In order to perform the simulations in the best-case scenario, we considered Ohmic contacts  by heavily $n$-doping the source and drain regions ($L_{\text{S/D}}$=2~nm) to reach a Fermi energy degeneracy of 0.2~eV,  with an equivalent electron doping concentration of 4.78$\cdot10^{14}$~m$^{-2}$ in these regions. In addition, the channel was electrostatically doped by a metallic back-gate whose voltage ($V_{\textbf{BG}}$) was set to $-$10~V unless explicitly stated otherwise. The BG is separated from the channel by a SiO$_2$ back insulator with equivalent thickness $t_{\text{bot}}$=10~nm $(\varepsilon_r\!=\!3.9)$. 
We embedded the CNT-gate in ZrO$_2$ $(\varepsilon_r\!=\!25.0)$ and placed it on top of the channel, i.e. on the opposite side of the back-gate with respect to the MoS$_2$, setting a distance $t_{\text{top}}$=5~nm between the center of the CNT and the channel. We modeled both the MoS$_2$ channel and the CNT through a semi-empirical nearest-neighbor tight-binding approximation: the Hamiltonian of the 2D material has been derived from a model for pseudo-hexagonal lattices, as in \cite{Tarun2015}. For the single-walled CNT we have considered a chiral vector guarantying semi-metallic states in the zig-zag configuration \cite{Wong2011}. The Non-Equilibrium Green's functions (NEGF) \cite{Datta2000a} are solved independently for the MoS$_2$ channel and the CNT, to obtain the charge density in each material which is later included in the Poisson equation to self-consistently determine the potential in the device.
For the sake of comparison, we also designed an alternative device structure, substituting the CNT with a metallic nanowire as top gate: here, differently from the previous scenario, the $V_{\text{G}}$ fixes the potential around the cylindrical gate. Lastly, the transmission coefficient (T) was determined considering 32 transversal modes to perform transport calculations. We then exploited T to compute the current along the MoS$_2$ channel following Landauer's approach \cite{Landauer} in the case of pure ballistic transport, fixing the source-to-drain voltage at $V_{\text{DS}}$=0.5~V.
\section{Simulation Results}
\begin{figure} [b]
	\vskip-5ex
	\centering
	\includegraphics[width=1.0\columnwidth]{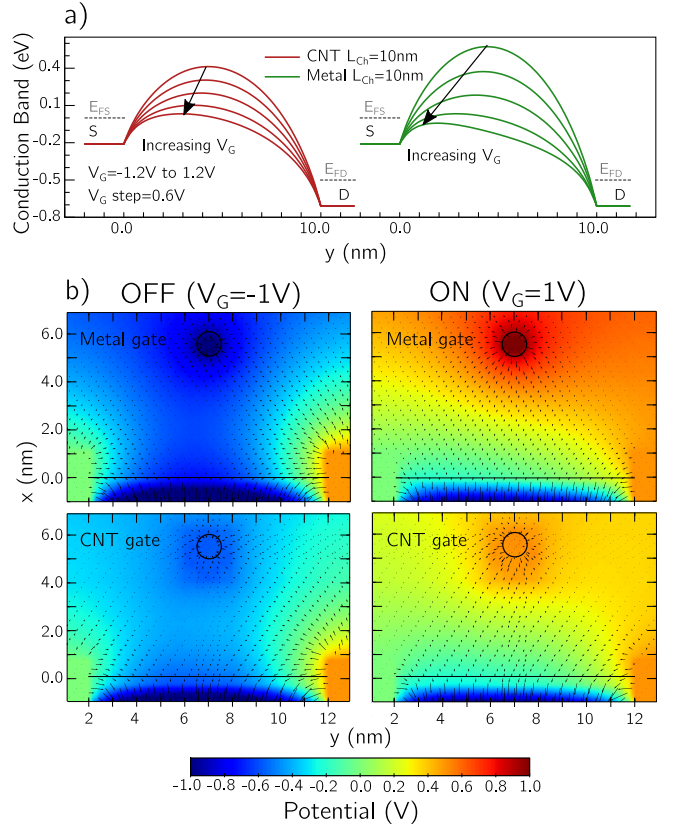}
	%\vskip-3ex
	\caption{(a) Conduction band profiles as function of top-gate voltage for CNT-gated (left picture) and metal-gated devices (right picture) with $L_{\text{Ch}}$=10~nm. $E_{FS}$ and $E_{FD}$ indicate the Fermi level of the source and drain, respectively. (b) Potential energy and electric field lines along the device section for metal-gated (top) and CNT-gated channels (bottom) at $V_{\text{G}}$=-1.0~V (left) and $V_{\text{G}}$=1.0~V (right); $V_{\text{DS}}$=0.5~V. Black solid lines and circles  highlight the position of the channel and the gate respectively. The arrows indicate the direction and intensity of the electric field along the channel (y-axis) and thickness (x-axis) dimensions.}  
	\label{metallic comparison}
	%\vskip-5ex
\end{figure}  
In order to capture the switching behavior and the ultimate performance of a uniformly ultra-scaled device, we will proceed as follows: we will start by examining devices of two different channel lengths with either a CNT or a metallic cylindrical gate; in addition, for the latter configuration we will investigate on the effect of $L_{\text{Ch}}$ by separating the different current components, and we will conclude by proposing some optimization parameters. 
We extracted the $\!I-\!V$ characteristics for $L_{\text{Ch}}$=7~nm and $L_{\text{Ch}}$=10~nm gated with a nanotube of $D$=1~nm. The curves, shown in logarithmic and linear scale in Fig. \ref{Structure}b with dashed lines, report a $SS$ far from the ideal limit of 60~mV/dec reaching at best 300~mV/dec for the longest channel length. The ratio of the ON-current to the OFF-current is less than two orders of magnitude, which means that it does not comply with the specifications for high-performance applications defined by the International Roadmap for Semiconductors (ITRS)\cite{ITRS}. The $I_{\textbf{OFF}}$ values were taken at 10$^{-1}$~A/m and, when not possible, to the lowest value of the current in the considered voltage window, the $I_{\textbf{ON}}$ values were consequently extracted at $V_{\textbf{GS}}=V_{\textbf{OFF}}$+0.5~V.  To understand the extent to which the finite density of states (DOS) could hinder the $\!I-\!V$ characteristics, we replaced the CNT with a cylindrical ideal metallic gate with continuous DOS, as an upper-limit case benchmark. A first observation must be done on the base of the results of the transfer characteristics in Fig.\ref{Structure}b comparing simulations with the same $L_{\text{Ch}}$ and different gate types. If we focus on the longest channel, the $SS$ improvement with the metallic gate is evident where the $SS$ lowers from 300~mV/dec to 160~mV/dec; the $I_{\text{ON}}$/$I_{\text{OFF}}$ as well increases to 400. By looking at the conduction band (CB) profile in the channel as a function of the applied $V_{\text{G}}$ (Fig. \ref{metallic comparison}a), it is possible to notice a worse electrostatic control and less modulation of the carbon nanotube with respect to the  metal nanowire gate. Indeed, a $\Delta V_{\text{G}}$ of 2.4~V results in a 0.35~eV reduction of the top of the barrier in the former case, half of what is achieved for the latter (0.7~eV). This effect can be also verified by comparing the potential profile and electric field for the different structures. Figure \ref{metallic comparison}b depicts the colormap of the potential for both the OFF- (left) and ON-state (right) in the metal gated (top) and CNT-gated devices (bottom). It is clear from the colormap that already in close proximity of the gate, the effective potential and electric field are significantly reduced when using a CNT. A reasonable explanation can be found in the semi-metallic nature of the carbon nanotube, i.e. its lack of a bandgap and its low DOS: less charge on the gate translates into fewer carriers in the MoS$_2$ channel. In other words, the quantum capacitance $C_{\text{q}}$ of the CNT, added in series to the oxide capacitance $C_{\text{ox}}$, reduces the total gate capacitance $C_{\text{G}}$.
\begin{figure}[b]
	\vskip-5ex
	\centering
	\includegraphics[width=1\columnwidth]{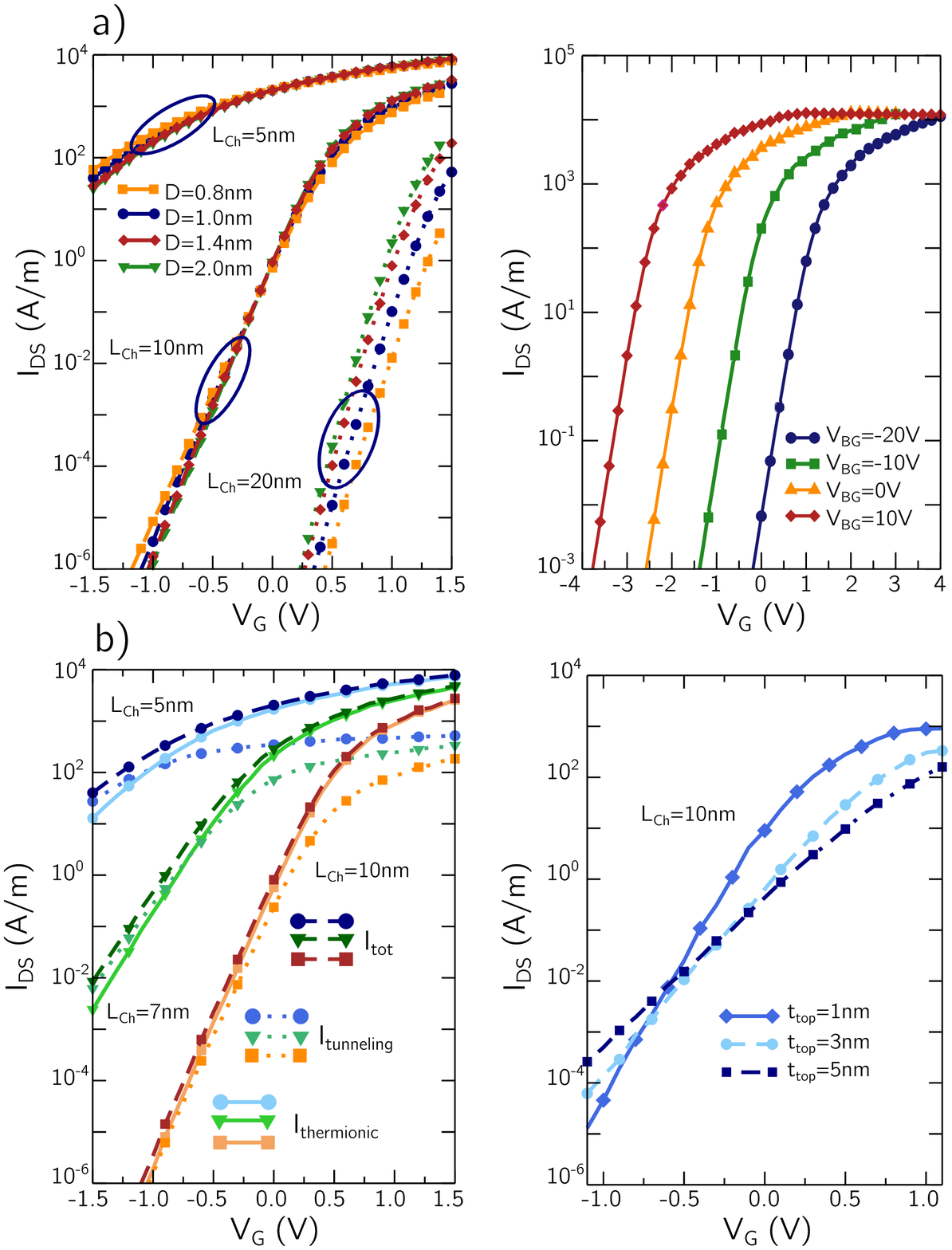}
	%\vskip-3ex
	\caption{(a) $\!I-\!V$ curves for metal-gated devices in semi-logarithmic scale as function of gate diameter and MoS$_2$ channel length with V$_{BG}$=$-$10~V (left panel) and as function of applied V$_{BG}$ for a 10~nm long device (right panel). (b) Total, thermionic and tunneling current components in semi-logarithmic scale for devices of different channel lengths with a cylindrical $D=$1~nm metal top-gate (left); $\!I-\!V$ characteristic as function of the top oxide (ZrO$_2$) thickness for a CNT-gated device with $L_{\text{Ch}}$=10~nm (right).}  
	\label{currentcomponents}
	\vskip-7ex	
\end{figure} 
If on the one hand materials properties can explain the limited channel control, on the other hand they alone cannot fully justify a $SS$ so far from the ideal. 
In Fig. \ref{currentcomponents}a the curves for various nanowire gate diameters (from $D$=0.8~nm to $D$=2.0~nm) are presented in semi-logarithmic scale, grouped as function of the different channel lengths ($L_{\text{Ch}}$=5~nm, 10~nm, 20~nm). Once the finite DOS limitation is removed, it becomes clear that the channel length is a major parameter in determining the device performance: the $SS$ worsens from 105~mV/dec for $L_{\text{Ch}}$=20~nm to 160~mV/dec for $L_{\text{Ch}}$=10~nm and finally to $\sim$~340~mV/dec for $L_{\text{Ch}}$=5~nm, whereas the influence of the diameter is almost completely negligible. For what concerns the backgate voltage, as can be seen in Fig. \ref{currentcomponents}a (right panel) it only leads to a rigid shift of the $\!I-\!V$ curves, as a result of the different induced charge density, while $SS$ is not altered. To gain a better insight into the non-ideal behavior of the MOSFETs, we  split the output current of metal-gated transistors in the thermionic and tunneling components (Fig. \ref{currentcomponents}b, left panel). The latter refers to electrons having energies lower than the top of the barrier which can tunnel through the barrier if this is sufficiently short: the smaller the $L_{\text{Ch}}$ (below 10~nm) the higher the tunneling current. In fact, strong SCEs limit both CNT-gated and metal-gated configurations, reasonably as a result of the greater influence of the contacts on the channel region which interferes with the gate control. Nonetheless, little improvements can be obtained in the overall behavior of the initial structure (Fig. \ref{Structure}a)  by carefully tuning the device electrostatics. For instance, we reduced the top ZrO$_2$ thickness from 5~nm to 3~nm and pushed it even to a technologically premature 1~nm. The right panel of Fig. \ref{currentcomponents}b reports the $I\!-\!V$ characteristics for the three different cases and the reduction in the $SS$s from 317~mV/dec, 256~mV/dec, to 172~mV/dec goes together with the decreasing of the oxide thickness. 

\section{Conclusion}
In this work we investigated the performance of field-effect transistors with MoS$_2$ channel and with a gate electrode consisting of a metallic CNT of diameter 1~nm via fully ballistic quantum transport simulations. We then analyzed the detrimental impact of the finite density of states of the nanotube by comparing the CNT-gated devices with those with an ideal metallic cylinder of the same dimension, representing an upper-limit case scenario. We were able to notice larger values of $SS$ for short-channel structures compared to the ideal 60~mV per decade of thermionic devices,  which negatively affect the switching behavior of the transistors. We attributed the poor gate control to SCEs such as the presence of tunneling currents and to gate-drain interactions. Finally, in order to optimize the initial CNT-gated configuration, we indicated few parameters to tune such as the top oxide thickness able to reduce the $SS$ and improve the $I\!-\!V$ characteristics, unlike the gate diameter and the back-gate voltage. 

\
\

The authors acknowledge the support from the Graphene Flagship (No. 785219). 
The authors gratefully acknowledge fruitful discussions with S.B. Desai and A. Javey
\

\nocite{*}
\bibliography{MoS2-CNT2_PROOF}% Produces the bibliography via BibTeX.

\end{document}